\documentclass[pdflatex,sn-nature]{sn-jnl}

\usepackage{graphicx}%
\usepackage{multirow}%
\usepackage{amsmath,amssymb,amsfonts}%
\usepackage{amsthm}%
\usepackage{mathrsfs}%
\usepackage[title]{appendix}%
\usepackage{xcolor}%
\usepackage{textcomp}%
\usepackage{manyfoot}%
\usepackage{booktabs}%
\usepackage{algorithm}%
\usepackage{algorithmicx}%
\usepackage{algpseudocode}%
\usepackage{listings}%

\theoremstyle{thmstyleone}%
\theoremstyle{thmstyletwo}%

\theoremstyle{thmstylethree}%

\begin{document}

\title[ADEPT–PolyGraphMT: Automated Molecular Simulation and Multi-Task Multi-Fidelity Machine Learning for Polymer Property Generation and Prediction]{ADEPT–PolyGraphMT: Automated Molecular Simulation and Multi-Task Multi-Fidelity Machine Learning for Polymer Property Generation and Prediction}


\author[1,2]{\fnm{Sobin} \sur{Alosious}}

\author[3]{\fnm{Yuhan} \sur{Liu}}

\author[2]{\fnm{Jiaxin} \sur{Xu}}

\author[4]{\fnm{Gang} \sur{Liu}}

\author[2]{\fnm{Renzheng} \sur{Zhang}}

\author[1,4]{\fnm{Meng} \sur{Jiang}}

\author*[1,2,5,6]{\fnm{Tengfei} \sur{Luo}}\email{tluo@nd.edu}


\affil[1]{\orgdiv{Lucy Family Institute for Data and Society}, 
\orgname{University of Notre Dame}, 
\orgaddress{\city{Notre Dame}, \state{IN}, \postcode{46556}, \country{USA}}}

\affil[2]{\orgdiv{Department of Aerospace and Mechanical Engineering}, 
\orgname{University of Notre Dame}, 
\orgaddress{\city{Notre Dame}, \state{IN}, \postcode{46556}, \country{USA}}}

\affil[3]{\orgdiv{Department of Chemistry and Biochemistry}, 
\orgname{University of Notre Dame}, 
\orgaddress{\city{Notre Dame}, \state{IN}, \postcode{46556}, \country{USA}}}

\affil[4]{\orgdiv{Department of Computer Science and Engineering}, 
\orgname{University of Notre Dame}, 
\orgaddress{\city{Notre Dame}, \state{IN}, \postcode{46556}, \country{USA}}}

\affil[5]{\orgdiv{Department of Chemical and Biomolecular Engineering}, 
\orgname{University of Notre Dame}, 
\orgaddress{\city{Notre Dame}, \state{IN}, \postcode{46556}, \country{USA}}}

\affil[6]{\orgdiv{Center for Sustainable Energy at Notre Dame (ND Energy)}, 
\orgname{University of Notre Dame}, 
\orgaddress{\city{Notre Dame}, \state{IN}, \postcode{46556}, \country{USA}}}


\abstract{The discovery of polymers with targeted properties is challenged by the vast chemical design space and the limited availability of consistent, high-quality data across multiple properties. In this work, an integrated polymer informatics framework is presented that combines the Automated molecular Dynamics Engine for Polymer simulaTions (ADEPT) workflow with multi-task and multi-fidelity machine learning (PolyGraphMT). Polymer repeat units are represented as molecular graphs and processed using a graph neural network to learn structure–property relationships. Starting from SMILES representations for monomers, ADEPT automates the construction of atomistic models and the evaluation of their properties using molecular dynamics simulations and density functional theory calculations. The simulation data are combined with curated experimental data and group contribution theory estimates to construct a unified dataset of approximately 62,000 polymer property values spanning 28 properties across thermal, mechanical, transport, electronic, optical, and structural properties. Using this dataset, inter-property correlations are analyzed, and multi-task learning strategies are evaluated for joint property prediction. The results show that multi-task models achieve performance comparable to single-task models in data-rich regimes and exhibit superior accuracy as training data become limited. In addition, fidelity-aware training improves predictive accuracy when combining experimental and computational data sources. The trained models are further applied to large-scale property prediction for polymers in the PolyInfo database ($\sim$13,000 polymers) and the PI1M virtual polymer library ($\sim$1 million polymers), producing physically consistent property distributions across a broad chemical space. Overall, the proposed framework provides a structured approach for scalable prediction and screening of polymer properties across multiple property types and data fidelity levels.}




\maketitle

\section{Introduction}
Polymers occupy a central role in modern materials science due to their exceptional chemical diversity and tunability across thermal, mechanical, transport, and electronic properties~\cite{bicerano2002prediction,sperling2015introduction}. This versatility underpins applications ranging from structural materials and membranes to energy storage, electronics, and biomedical devices~\cite{mike2013recent,bujak2013polymers,lyu2009degradability}. At the same time, identifying polymers with targeted combinations of properties remains a fundamental challenge, as the accessible chemical space is effectively unbounded and experimental characterization is both time-consuming and property-specific~\cite{audus2017polymer,otsuka2011polyinfo}.

Data availability is a central requirement for materials informatics, as machine learning (ML) models rely critically on the quality, quantity, and diversity of training data~\cite{ramprasad2017machine,alosious2025computation}. In inorganic materials science and small-molecule chemistry, large open databases derived from high-throughput first-principles calculations and curated experimental measurements, such as the Materials Project, AFLOW, OQMD, and QM9, have played a decisive role in advancing data-driven materials discovery \cite{jain2013commentary,curtarolo2012aflow,saal2013materials,ramakrishnan2014quantum}. In the polymer domain, dedicated databases such as PolyInfo and Polymer Genome have enabled important progress by compiling experimental measurements and first-principles data for selected properties \cite{otsuka2011polyinfo,kim2018polymer}. While these databases have enabled important progress in polymer informatics, the underlying data remain highly sparse and heterogeneous. For most polymers, only a limited number of properties are reported, often measured under different processing histories, molecular weights, and testing conditions.
In addition, recent studies have demonstrated the feasibility of high-throughput molecular dynamics (MD) workflows for generating polymer property data at scale~\cite{afzal2020high,hayashi2022radonpy}. However, these efforts typically target a narrow set of properties and datasets, ranging from hundreds to a few thousand polymers, and are often generated using study-specific protocols. As a result, polymer property data remain fragmented, with limited multi-property overlap and weak integration across experimental and computational fidelities, and have not yet achieved the scale or uniformity of inorganic and small-molecule databases. More recently, large-scale efforts have further extended high-throughput polymer simulation workflows toward the construction of comprehensive data resources. Yoshida et al.~\cite{yoshida2025omics} introduced the PolyOmics database, an omics-scale computational polymer dataset generated using fully automated MD pipelines. The database comprises physical property data for over $10^5$ polymeric materials and includes a diverse set of thermal, mechanical, dielectric, and transport properties computed under standardized simulation protocols. In contrast to earlier datasets limited in size and scope, PolyOmics demonstrates that large-scale simulation data can be effectively leveraged for machine learning through simulation-to-real (Sim2Real) transfer learning, where models pretrained on computational data are fine-tuned using limited experimental measurements. The study further reports power-law scaling behavior in prediction accuracy with increasing dataset size, highlighting the potential of ultralarge computational datasets as foundational resources for polymer informatics.

At the same time, it is well established that simulation-derived properties, particularly those obtained from classical MD, exhibit systematic deviations from experimental measurements due to force-field limitations and the classical treatment of atomic vibrations~\cite{mcquarrie2008quantum,jenniferlukes}. Despite these systematic offsets, MD predictions consistently preserve strong correlations with experimental trends for multiple polymer properties, including thermal conductivity ($\kappa$), specific heat capacity ($C_p$), density ($\rho$), and glass transition temperature ($T_g$), across chemically diverse polymer classes~\cite{zhang2025active,xu2024unlocking,ma2022machine,buchholz2002cooling}. This behavior indicates that simulation data encode meaningful structure–property relationships and are well-suited as lower-fidelity proxies for learning global trends, rather than as direct substitutes for experimental measurements.

Several studies have focused on developing automated workflows for polymer simulation to address data scarcity. Hayashi et al.~\cite{hayashi2022radonpy} introduced RadonPy, an open-source Python framework that enables fully automated, high-throughput all-atom MD simulations for predicting polymer properties. The workflow integrates polymer structure generation from SMILES representations, force-field assignment, equilibration, equilibrium and nonequilibrium MD simulations, and post-processing of trajectories within a unified pipeline, enabling systematic MD data generation directly from chemical structure. Complementing this effort, Turney and Matta~\cite{turney2025atomistic} presented a perspective on atomistic polymer modeling that emphasizes persistent bottlenecks in system building and force-field parametrization workflows. Their review highlights the lack of standardized, end-to-end pipelines capable of robustly handling polymer-specific features such as tacticity, branching, and polydispersity, and surveys 11 open-source polymer-building tools with respect to functionality, maintenance status, and architectural support. Together, these studies underscore both the progress and remaining challenges in scalable polymer MD workflows.

ML has emerged as a powerful framework for predicting polymer properties directly from chemical structure, enabling rapid evaluation of materials that would be infeasible to characterize experimentally or computationally at scale~\cite{ramprasad2017machine}. Single-task learning approaches, in which independent models are trained for individual properties, have demonstrated promising accuracy for targets such as $T_g$, elastic moduli, and dielectric response~\cite{mannodi2016machine}. However, these models are inherently data-inefficient for sparsely sampled properties and do not exploit the strong correlations among polymer properties~\cite{sha2021machine}.

Multi-task learning addresses these limitations by learning shared representations across multiple tasks, enabling information transfer between related properties. Kuenneth et al.~\cite{kuenneth2021polymer} demonstrated that multi-task neural networks trained on a dataset comprising 36 polymer properties for approximately 13,000 polymers consistently outperform single-task models, with error reductions of up to 20--30\% for sparsely sampled properties. Similarly, Queen et al.~\cite{queen2023polymer} developed POLYMERGNN, a graph neural network framework for multi-task prediction using an experimentally curated dataset of over 240 synthesized polyesters, achieving predictive performance of $R^2 \approx 0.72$ for $T_g$ and $R^2 \approx 0.70$ for intrinsic viscosity under five-fold cross-validation. Although this work focused exclusively on experimental data, it demonstrated the effectiveness of shared latent representations for learning correlated polymer properties.

Despite these advances, polymer property datasets remain inherently heterogeneous, combining experimental measurements with MD, density functional theory (DFT), and group contribution (GC) data that differ substantially in fidelity. Treating all data sources as equally reliable can bias learned models toward abundant but lower-fidelity data, while discarding computational data limits chemical coverage. Although multi-fidelity learning strategies have been explored in broader materials science contexts, their systematic integration with multi-task learning for polymer property prediction remains limited.~\cite{pilania2017multi,wang2021compositionally,liu2025machine}

In this work, we introduce ADEPT, an Automated molecular Dynamics Engine for Polymer simulaTions, for systematic generation of polymer properties from chemical structure, and integrate it with a newly developed multi-task, multi-fidelity ML framework (referred to here as PolyGraphMT) for large-scale polymer property prediction. ADEPT is designed for high-performance computing, enabling massively parallel, high-throughput property calculations using property-specific MD workflows, including both equilibrium and nonequilibrium methods. It combines atomistic MD and DFT calculations with curated experimental data and GC estimates to construct a heterogeneous dataset of approximately 62{,}000 data points spanning 28 properties across thermal, mechanical, transport, electronic, optical, and structural domains.
Using this dataset, we analyze inter-property correlations, evaluate task grouping strategies, and quantify the benefits of multi-task learning under varying data availability while accounting for data fidelity. The PolyGraphMT framework represents polymer repeat units as molecular graphs and employs graph neural networks to learn structure--property relationships across multiple properties and fidelity levels.
The integrated ADEPT–PolyGraphMT framework enables (i) simultaneous multi-property prediction, (ii) fidelity-aware learning across heterogeneous data sources, and (iii) scalable screening over large polymer spaces.
The resulting models enable unified prediction and consistent screening across both experimental polymer databases ($\sim 13{,}000$ polymers) and large virtual polymer libraries ($\sim 1$ million polymers), which is not achievable with existing fragmented workflows.
This establishes a scalable framework for data-driven polymer discovery by integrating property generation, model training, and large-scale screening.
The overall framework integrates the ADEPT simulation engine with the PolyGraphMT learning architecture.

\section{Methodology}
\subsection{ADEPT Workflow}

\begin{figure}[b!]
  \centering
  \includegraphics[width=1.0\textwidth]{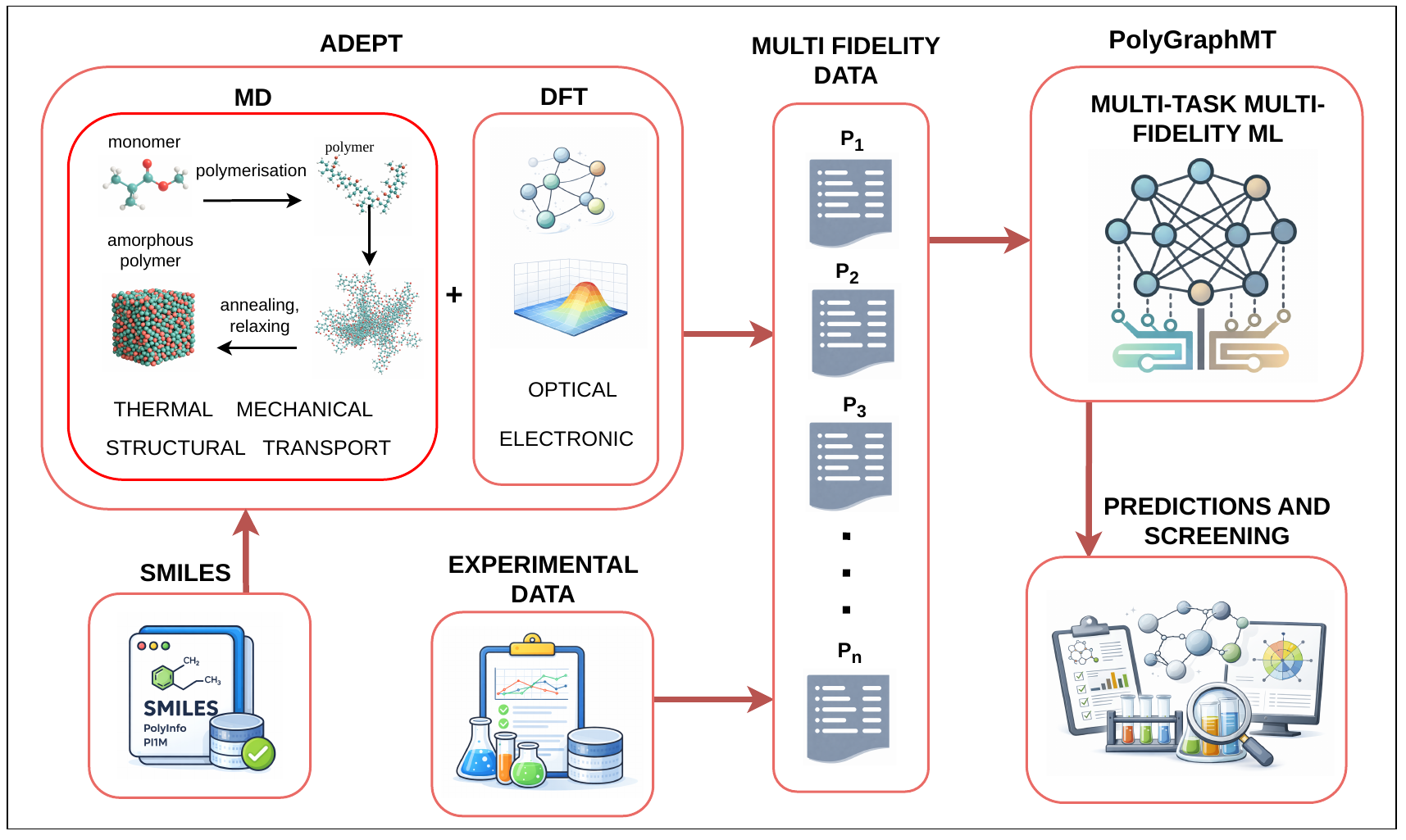}
  \caption{Overview of the ADEPT and PolyGraphMT workflow. Polymer SMILES are processed through the ADEPT pipeline to generate thermal, mechanical, structural, transport, electronic, and optical properties using MD and DFT. These computational data are combined with curated experimental measurements to construct a unified dataset, which is used to train multi-task, multi-fidelity ML models for polymer property prediction and large-scale screening.}
  \label{schematic}
\end{figure}

Figure~\ref{schematic} presents an overview of the end-to-end methodology developed in this work. The workflow begins with polymer chemical structures represented as SMILES strings, which provide a common input representation for all subsequent stages. These structures are processed using the ADEPT framework to automatically generate atomistic polymer models, perform equilibration and production simulations, and compute a broad range of polymer properties using all-atom MD. DFT calculations are additionally employed to obtain selected electronic and optical properties that are not  accessible from classical molecular simulations. To improve clarity, detailed methodologies, including polymer structure generation from SMILES, amorphous polymer construction and equilibration protocols, property calculations using MD simulations, and electronic and dielectric property evaluations from DFT, are provided in the Methods section (see Methods – ADEPT Workflow: Polymer Simulation and Property Generation). The simulation-derived properties are combined with experimentally measured data curated from the literature to construct a unified dataset spanning thermal, mechanical, transport, electronic, optical, and structural property domains. Due to the different origins of the data, the resulting dataset contains multiple levels of data fidelity, with experimental measurements treated as the highest fidelity, followed by DFT, MD simulations, and GC estimates. While experimental data from the literature will undoubtedly contain noise depending on the source of the data, they are the best reflection of true polymers in the real world. Any computational models have certain levels of approximations. This dataset is used to train multi-task, multi-fidelity ML models that learn shared chemical representations across properties while accounting for systematic differences between experimental and computational data sources.  The trained models are then applied to property prediction and large-scale screening of both experimentally reported polymers and virtual polymer libraries, enabling efficient exploration of polymer chemical space. The ADEPT workflow and the associated ML pipelines are released as open-source software, as described in the Data and Code Availability section.

\subsection{PolyGraphMT Workflow}

\subsubsection{Learning objective and data structure}

To enable scalable prediction of polymer properties beyond direct molecular simulations, a multi-task, multi-fidelity ML framework was developed using data generated by the ADEPT workflow, along with complementary experimental and GC datasets. The objective of the model is to jointly learn multiple polymer properties while accounting for differences in data fidelity arising from experimental measurements, MD simulations, DFT calculations, and GC estimates. The implementation builds upon and substantially extends the \texttt{torch-molecule} framework~\cite{liu2022graph}, which was originally developed for single-task molecular property prediction.

Each polymer is represented by its repeat-unit chemical structure and may be associated with one or more target properties evaluated at different fidelity levels. Let $i \in \{1,\dots,N\}$ index polymers, $p \in \mathcal{P}$ index target properties (for example, $C_p$, $\kappa$, and $T_g$), and $f \in \mathcal{F}$ index data fidelity levels corresponding to experimental, MD, DFT, or GC sources.

For a given polymer $i$, property $p$, and fidelity level $f$, the observed value is denoted as $y_{i,p}^{(f)}$. In practice, only a subset of all possible $(i,p,f)$ combinations is available, leading to a sparse, heterogeneous supervision structure. The learning objective is to construct a predictive model that leverages shared information across related properties and fidelity levels, while treating experimental data as the highest-fidelity reference during model training and evaluation.

\subsubsection{Molecular graph representation}

Polymer repeat units were represented as molecular graphs constructed from SMILES strings. For each polymer $i$, the molecular graph is defined as $\mathcal{G}_i = (\mathcal{V}_i, \mathcal{E}_i)$, where $\mathcal{V}_i$ denotes the set of atoms (nodes) and $\mathcal{E}_i$ denotes the set of covalent bonds (edges). Node features encode atomic identity and local chemical attributes, while edge features represent bond connectivity and bond order.

This graph-based representation preserves the chemical structure of the repeat unit and enables the use of message-passing neural networks to learn structure–property relationships directly from molecular connectivity \cite{gilmer2017neural}.

\subsubsection{Graph neural network encoder}

A shared graph neural network encoder was used to map each polymer graph $\mathcal{G}_i$ to a fixed-dimensional latent representation $\mathbf{z}_i \in \mathbb{R}^d$. This mapping is expressed as
\begin{equation}
\mathbf{z}_i = f_{\boldsymbol{\theta}}(\mathcal{G}_i),
\end{equation}
where $f_{\boldsymbol{\theta}}$ denotes the encoder parameterized by weights $\boldsymbol{\theta}$.

The encoder comprises multiple graph convolution layers in which atomic feature vectors are iteratively updated through neighborhood aggregation. At each layer, atomic representations are updated using information from directly bonded neighbors, followed by nonlinear activation and normalization. After message passing, a global pooling operation aggregates the node-level representations into a single polymer-level embedding $\mathbf{z}_i$. This shared embedding serves as the common input for all downstream property prediction tasks and follows standard message-passing neural network implementations \cite{gilmer2017neural}. Additional implementation details of the network architecture and training configuration are provided in the Methods section (see Methods – PolyGraphMT Framework).

\subsubsection{Property-specific prediction heads}

For each target property $p$, a dedicated prediction head maps the shared latent representation $\mathbf{z}_i$ to a scalar prediction,
\begin{equation}
\hat{y}_{i,p} = g_p(\mathbf{z}_i),
\end{equation}
where $g_p$ denotes a property-specific multilayer perceptron with parameters that are independent across properties.

All prediction heads operate on the same encoder output $\mathbf{z}_i$, which enforces a shared representation across tasks while allowing property-dependent nonlinear transformations in the final layers. This design supports information sharing among related properties without requiring identical functional forms for different prediction targets \cite{caruana1997multitask}.

\subsubsection{Multi-fidelity supervision strategy}

When a target property $p$ is available at multiple fidelity levels for a given polymer, all corresponding observations are included during training. The model produces a single property prediction $\hat{y}_{i,p}$ from the shared latent representation, and this prediction is compared independently against each available fidelity-specific observation $y_{i,p}^{(f)}$ through the training loss.
This formulation incorporates heterogeneous data sources without introducing separate prediction branches or explicit bias-correction terms for individual fidelity levels. Differences in data fidelity are handled through loss weighting during optimization, allowing lower-fidelity data to contribute to representation learning while higher-fidelity experimental data exert a stronger influence during model fitting and evaluation. Detailed formulations of the loss function, normalization strategy, and training procedures are provided in the Methods section (see Methods – PolyGraphMT Framework: Multi-Task Multi-Fidelity Learning).

\section{Results and Discussions}

\begin{table*}[h!]
\centering
\caption{Summary of polymer properties used in this work, including symbols, units, data sources, number of data points, and corresponding value ranges.}
\label{table1}
\begin{tabular}{lccccl}
\hline
\textbf{Property} & \textbf{Symbol} & \textbf{Unit} & \textbf{Source} & \textbf{Points} & \textbf{Data range} \\
\hline
\multicolumn{6}{l}{\textbf{Thermal}} \\
Melting temperature                & $T_m$      & K             & Exp.   & 3671 & [210.6--873.1] \\
Glass transition temperature       & $T_g$      & K             & Exp./MD       & 7360 & [134.1--768.1] \\
Thermal diffusivity               & $\alpha_{\rm T}$ & m$^2$/s & MD & 799 & [3.5$\times10^{-8}$--8.3$\times10^{-7}$] \\
Thermal conductivity               & $\kappa$ & W/m.K & MD/Exp. & 2327 & [0.002--1.59] \\
Specific heat capacity             & $C_p$ & J/ kg.K & Exp./MD/GC & 13105 & [439.2--2831.7] \\[4pt]

\multicolumn{6}{l}{\textbf{Mechanical}} \\
Young’s modulus                    & $E$        & GPa          & MD         & 1012 & [0.36--11.04] \\
Shear modulus                      & $G$        & GPa          & MD         & 1012 & [0.12--4.2] \\
Bulk modulus                       & $K$        & GPa          & MD         & 1017 & [0.72--10.9] \\
Poisson ratio                      & $\nu$      & --           & MD         & 1012 & [-0.17--0.48] \\[4pt]

\multicolumn{6}{l}{\textbf{Transport}} \\
Viscosity                          & $\eta$     & Pa$\cdot$s   & MD         & 704 & [$9.5\times10^{-5}$--0.11] \\
Diffusivity                        & $D$        & cm$^2$/s & MD       & 700 & [$2.19\times10^{-9}$--0.096] \\[4pt]

\multicolumn{6}{l}{\textbf{Gas Permeability}} \\
He permeability                    & $P_\mathrm{He}$   & Barrer     & Exp.   & 466 & [0.05--17800] \\
H$_2$ permeability                 & $P_\mathrm{H_2}$  & Barrer     & Exp.   & 511 & [0.02--36800] \\
CO$_2$ permeability                & $P_\mathrm{CO_2}$ & Barrer     & Exp.   & 756 & [1.2$\times10^{-6}$--47000] \\
N$_2$ permeability                 & $P_\mathrm{N_2}$  & Barrer     & Exp.   & 798 & [1.6$\times10^{-4}$--16600] \\
O$_2$ permeability                 & $P_\mathrm{O_2}$  & Barrer     & Exp.   & 807 & [7.0$\times10^{-7}$--18700] \\
CH$_4$ permeability                & $P_\mathrm{CH_4}$ & Barrer     & Exp.   & 683 & [4.1$\times10^{-4}$--35000] \\[4pt]

\multicolumn{6}{l}{\textbf{Electronic / Optical}} \\
Polarizability                     & $\alpha$    & a.u.         & DFT        & 2036 & [1.9--70.3] \\
HOMO energy                        & $E_{\mathrm{HOMO}}$ & eV       & DFT        & 2916 & [-13.4-- -6.84] \\
LUMO energy                        & $E_{\mathrm{LUMO}}$ & eV       & DFT        & 2916 & [-3.01--3.08] \\
Band gap                           & $E_g$       & eV           & DFT        & 2916 & [5.4--16.7] \\
Dipole moment                      & $\mu$       & Debye        & DFT        & 2916 & [0.003--12.49] \\
Total electronic energy            & $E_{\mathrm{total}}$ & eV     & DFT        & 2916 & [-3.2$\times10^{-6}$-- -2.5$\times10^{-6}$] \\
Refractive index                   & $n$         & --           & MD+DFT      & 744 & [1.02--1.84] \\
Dielectric constant                & $\varepsilon$ & --         & MD+DFT         & 744 & [1.06--12.7] \\
Permittivity                       & $\epsilon_r$ & --          & MD+DFT     & 744 & [9.4--113.1] \\[4pt]

\multicolumn{6}{l}{\textbf{Structural / Physical}} \\
Radius of gyration                & $R_g$        &  \AA     & MD        & 2500 & [9.17--75.22] \\
Density                           & $\rho$       & g/cm$^{3}$  & MD        & 3644 & [0.11--2.97] \\
\hline
\end{tabular}
\end{table*}

The results presented in this section are obtained using the ADEPT workflow, which enables end-to-end evaluation of polymer properties starting from repeat-unit SMILES representations~\cite{liu2025open}. For a given polymer structure, ADEPT automates polymer construction, force-field assignment, MD simulations, electronic-structure calculations, and data processing to generate a wide range of thermodynamic, mechanical, transport, electronic, dielectric, and structural properties in a consistent and scalable manner.
Data generated using ADEPT, including MD and DFT results, are combined with experimentally measured values and GC estimates to construct a unified dataset of polymer properties. This integrated dataset spans multiple property classes and data fidelity levels, providing a physically meaningful and chemically consistent foundation for data-driven modeling. The resulting framework enables systematic validation of simulation-derived properties against experimental data and supports the development of multitask and multifidelity learning models based on a common structural representation. The following sections first evaluate the accuracy of MD-predicted properties and then examine how the combined experimental, simulation, and GC datasets are used to predict polymer properties using ML.

Table~\ref{table1} summarizes the polymer property dataset used in this work, comprising approximately 62,000 data points spanning thermal, mechanical, transport, gas permeability, electronic/optical, and structural properties. The dataset integrates information from multiple sources, including experiments, MD, DFT, and GC estimates, and covers a wide range of property magnitudes with highly variable data availability across tasks.
The thermal subset includes $T_m$, $T_g$, $\alpha_{\mathrm{T}}$, $\kappa$, and $C_p$, with $C_p$ and $T_g$ having the most extensive coverage and the most diverse fidelity composition. Mechanical properties ($E$, $G$, $K$, and $\nu$) and transport properties ($\eta$ and $D$) are obtained from MD simulations. Gas permeability properties ($P_{\mathrm{He}}$, $P_{\mathrm{H_2}}$, $P_{\mathrm{CO_2}}$, $P_{\mathrm{N_2}}$, $P_{\mathrm{O_2}}$, and $P_{\mathrm{CH_4}}$) are derived from experiments and span multiple orders of magnitude. Electronic and optical properties computed at the DFT level include $\alpha$, $E_{\mathrm{HOMO}}$, $E_{\mathrm{LUMO}}$, $E_g$, $\mu$, and $E_{\mathrm{total}}$, which are combined with MD-derived information to estimate bulk properties ($n$, $\varepsilon$, and $\epsilon_r$). Structural and physical properties ($R_g$ and $\rho$) provide direct links between polymer conformational statistics, packing behavior, and macroscopic properties.
Overall, the large number of target properties, heterogeneous data fidelity, and strongly imbalanced data availability across properties (Table~\ref{table1}) motivate the adoption of a unified multi-task, multi-fidelity ML framework.

\begin{figure}[t!]
  \centering
  \includegraphics[width=1.0\textwidth]{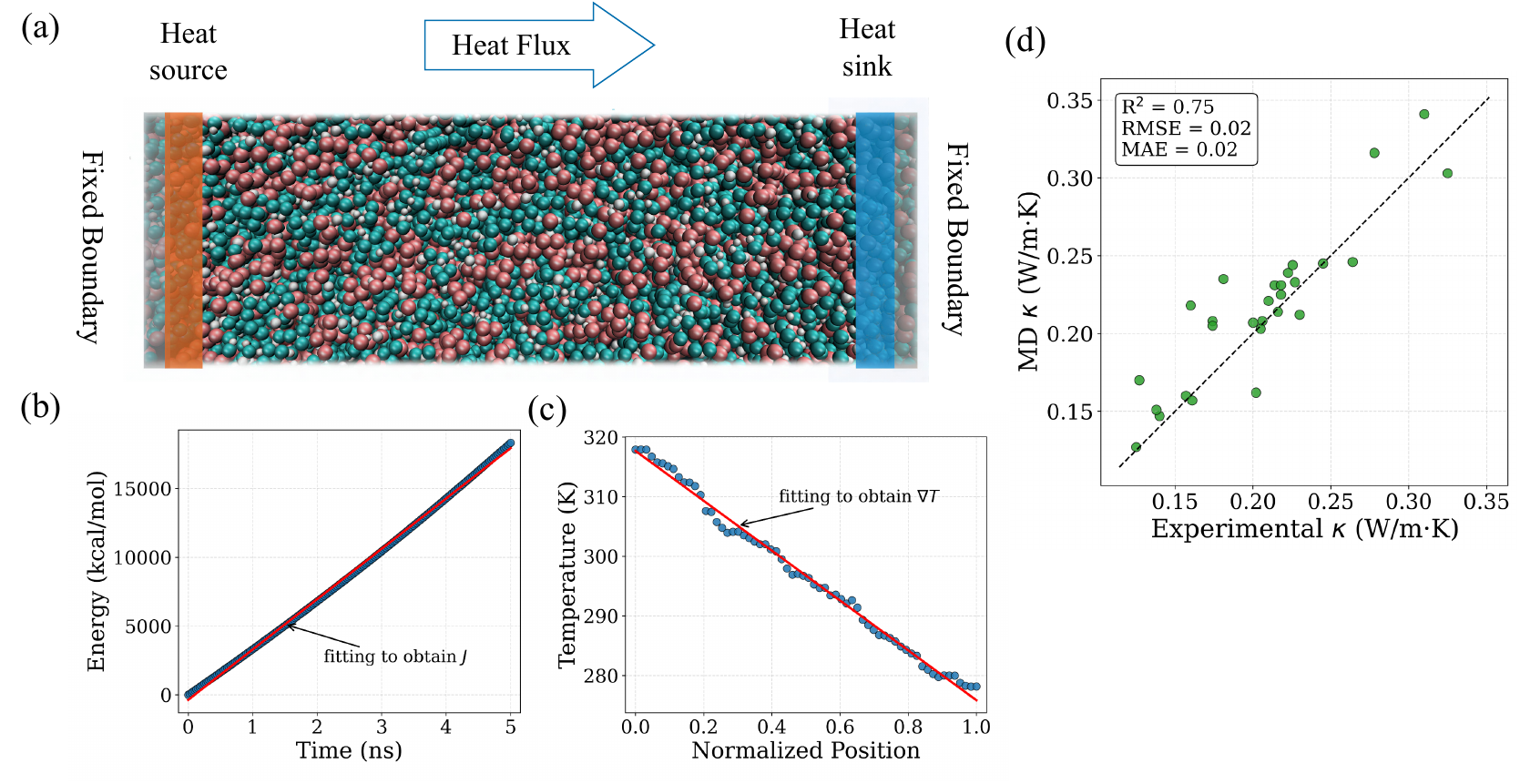}
  \caption{NEMD workflow and validation for $\kappa$ calculations.
(a) Representative MD snapshot illustrating the NEMD setup, where a constant heat flux is imposed between hot and cold regions using thermostatted slabs, with boundary layers composed of fixed atoms.
(b) Cumulative energy exchanged with the heat reservoirs as a function of time; a linear fit is used to obtain the heat flux $J$.
(c) Steady-state temperature profile along the transport direction; the linear region is fitted to extract the temperature gradient $\nabla T$.
(d) Comparison of $\kappa$ values obtained from MD simulations with experimental data, demonstrating good agreement between NEMD predictions and experiments.}
  \label{tc}
\end{figure}

\begin{figure}[t!]
  \centering
  \includegraphics[width=0.9\textwidth]{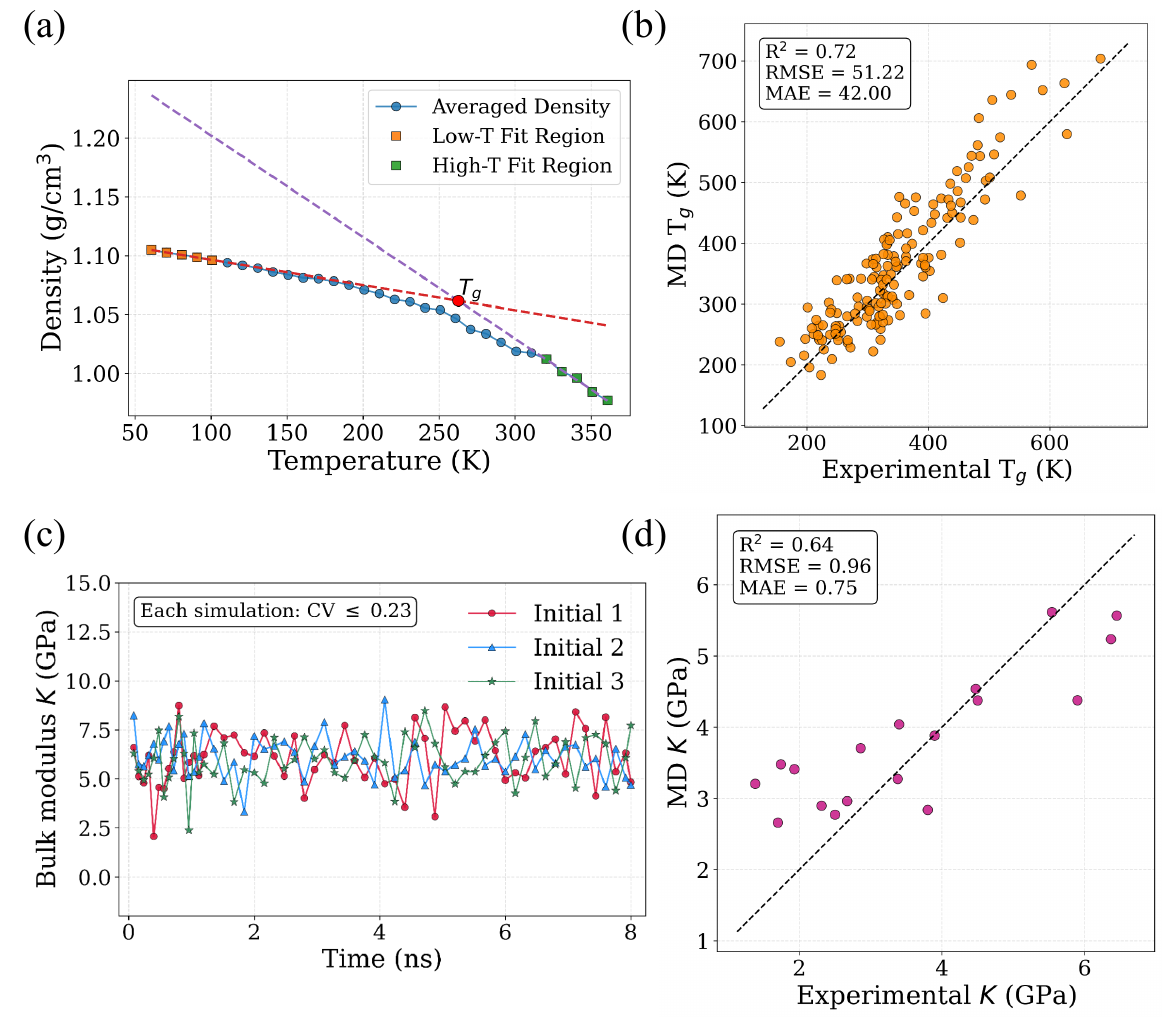}
  \caption{ 
(a) Temperature-dependent density profile used to estimate the  $T_g$ from the intersection of linear fits to the low- and high-temperature regimes. 
(b) Parity plot comparing MD-predicted and experimental $T_g$. 
(c) Time evolution of $K$ obtained from independent simulations with different initial configurations, illustrating temporal fluctuations and convergence. 
(d) Parity plot comparing MD-predicted and experimental  $K$.
}
  \label{tgmod}
\end{figure}

Figure~\ref{tc} presents the NEMD results used to validate the MD-derived $\kappa$ values used in this work. A representative simulation snapshot showing the imposed heat-flow direction and boundary configuration is presented in Fig.~\ref{tc}(a), confirming the formation of a well-defined transport geometry without structural distortion~\cite{ma2022machine}.
The cumulative energy exchanged with the heat reservoirs shows a clear linear dependence on time (Fig.~\ref{tc}(b)), indicating the establishment of a stable steady-state regime from which the heat flux is reliably extracted. The corresponding temperature profile along the transport direction exhibits a well-defined linear region within the central conduction zone (Fig.~\ref{tc}(c)), with minimal curvature and noise, enabling robust estimation of the temperature gradient.
The resulting $\kappa$ values are directly compared with experimental measurements in Fig.~\ref{tc}(d), considering only data reported between 20~$^\circ$C and 30~$^\circ$C to match the room-temperature MD simulation conditions. The MD predictions show good agreement with experiments, yielding an $R^2$ value of 0.75 and low absolute error. These results indicate that the MD workflow captures the dominant thermal transport trends across diverse polymer chemistries and provides physically consistent $\kappa$ values suitable for subsequent ML analysis.

Figure~\ref{tgmod}(a--b) illustrates the MD-based evaluation of $T_g$. In MD simulations, $T_g$ is identified from the temperature dependence of $\rho$ and reflects changes in molecular mobility and thermal expansivity in amorphous polymers.
Figure~\ref{tgmod}(a) shows the variation of $\rho$ with temperature for a representative polymer. Two approximately linear regimes are observed, corresponding to the glassy and rubbery states. To estimate $T_g$, linear fits are performed separately for the low-temperature and high-temperature regions using different numbers of data points. Multiple fits are performed, and the final $T_g$ is taken as the mean of the resulting intersection temperatures. If any fitted $T_g$ value falls outside the initial temperature-scanning range, it is discarded. This approach captures changes in volumetric thermal expansion while reducing sensitivity to the specific choice of fitting window. The parity plot in Fig.~\ref{tgmod}(b) compares MD-predicted $T_g$ values with experimental data and shows good agreement, with an $R^2$ value of 0.72 and low prediction error across chemically diverse polymers.

\begin{figure}[h!]
  \centering
  \includegraphics[width=0.77\textwidth]{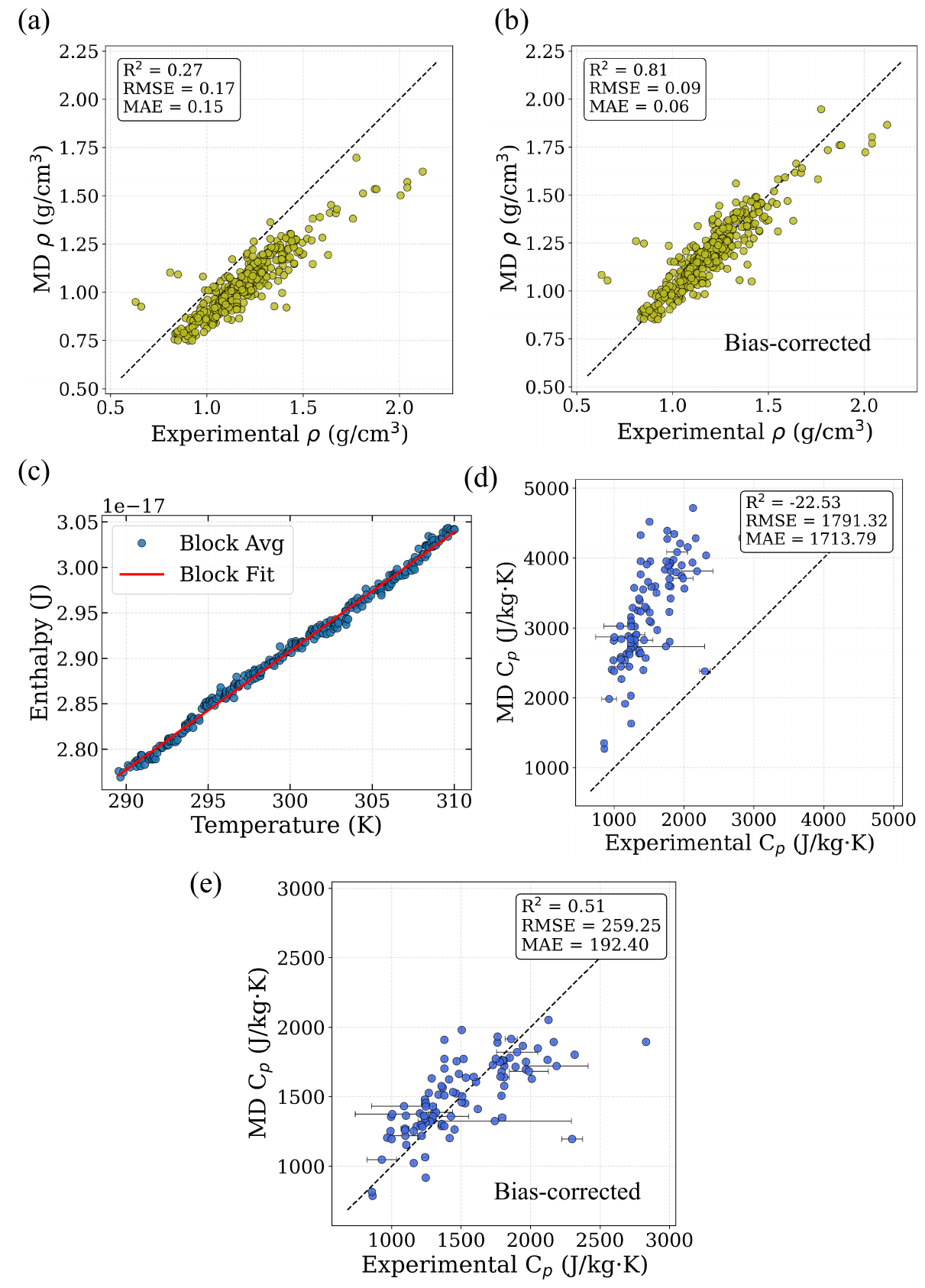}
  \caption{
(a) Parity plot comparing MD-predicted  $\rho$ with experimental values, showing systematic deviation.
(b) Density parity plot after bias correction, demonstrating improved agreement between MD and experimental $\rho$.
(c) Representative enthalpy–temperature relationship obtained from MD simulations; block-averaged enthalpy values are fitted linearly to extract $C_p$.
(d) Parity plot comparing MD-predicted $C_p$ with experimental $C_p$, highlighting significant bias in raw MD predictions.
(e) $C_p$ parity plot after bias correction, showing improved correlation and reduced error relative to experimental data.}
  \label{rhocp}
\end{figure}

Figure~\ref{tgmod}(c--d) presents the MD-based evaluation of the bulk modulus $K$. Figure~\ref{tgmod}(c) shows the time evolution of $K$ obtained from simulations initiated from different configurations, illustrating temporal fluctuations around a stable mean. To validate the reliability of the MD predictions, polymers with experimentally reported $K$ values were collected from PoLyInfo and other literature sources~\cite{wypych2022handbook}, considering only measurements conducted between 20~$^\circ$C and 30~$^\circ$C to match the room-temperature simulation conditions. The parity plot in Fig.~\ref{tgmod}(d) compares MD-predicted and experimental $K$ values and shows reasonable agreement, with an $R^2$ value of 0.64 and low absolute error.
The remaining deviations are attributed to variations in experimental conditions, polymer synthesis routes, and measurement techniques, which introduce noise into the reported modulus values. To assess the influence of polymer morphology on modulus evaluation, multiple independent initial structures were examined. The resulting variations in $K$ were small, indicating that modulus predictions are largely insensitive to morphological randomness in the initial configurations~\cite{liu2026activelearningenabledmultiobjectivedesign}. To further reduce potential morphology-related bias, all reported MD-derived $K$ values were averaged over simulations performed using three distinct initial structures for each polymer.

Figure~\ref{rhocp}(a--b) evaluates the accuracy of MD-predicted  $\rho$ through direct comparison with experimental data and illustrates the effect of bias correction. The parity plot in Fig.~\ref{rhocp}(a) shows a systematic deviation between MD predictions and experiments, despite a clear linear correlation across the sampled density range, indicating the presence of a global bias. The initial MD predictions yield a low $R^2$ value of 0.21. After applying a linear calibration using polymers common to both datasets, the agreement improves substantially, as shown in Fig.~\ref{rhocp}(b), with the $R^2$ increasing to 0.81. This corresponds to an improvement of approximately 286\%, demonstrating a significant enhancement in agreement with experimental $\rho$ values.

Figure~\ref{rhocp}(c--e) presents the corresponding analysis for the $C_p$. Figure~\ref{rhocp}(c) shows a representative enthalpy--temperature relationship obtained from MD simulations, where block-averaged enthalpy varies linearly with temperature, allowing stable estimation of $C_p$ from the slope. Direct comparison with experimental values reveals that MD-predicted $C_p$ is systematically overestimated, as shown in Fig.~\ref{rhocp}(d), consistent with previous studies based on classical MD simulations~\cite{hayashi2022radonpy,bhowmik2019calculation,alosious2026transfer}.
The overestimation of $C_p$ arises from the classical treatment of vibrational degrees of freedom in MD simulations. In classical mechanics, all vibrational modes are fully excited according to the Boltzmann distribution, independent of vibrational frequency. In contrast, quantum mechanics predicts partial population of high-frequency vibrational modes at room temperature, following Bose-Einstein statistics. As a result, classical MD assigns excess vibrational energy to stiff bond-stretching and bond-bending modes, leading to inflated $C_p$ values relative to experiments~\cite{mcquarrie2008quantum,jenniferlukes}.
Despite this systematic bias, MD-predicted $C_p$ values exhibit a strong monotonic correlation with experimental data, indicating that MD captures the underlying structure--property trends. This behavior and its implications for data-driven correction and transfer learning have been analyzed in detail in our previous work~\cite{alosious2026transfer}. The initial MAE of 1713.79~J/kg.K is reduced to 192.40~J/kg.K after applying a global linear correction, corresponding to an error reduction of approximately 89\%. The corrected parity plot in Fig.~\ref{rhocp}(e) shows improved agreement with experiments, with reduced scatter and substantially lower prediction error.

\begin{figure}[h!]
  \centering
  \includegraphics[width=1.05\textwidth]{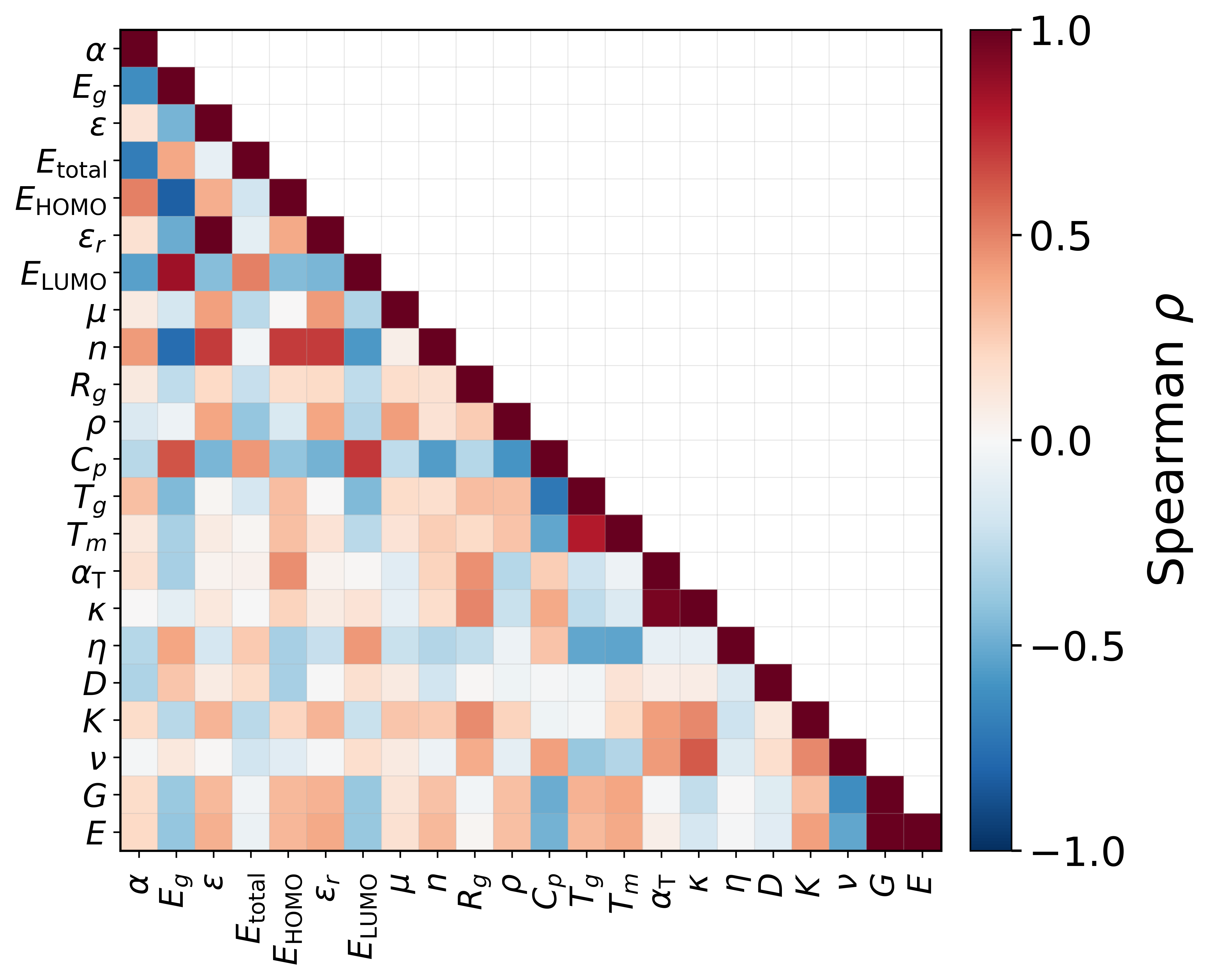}
  \caption{Polymer property heatmap of Pearson correlation coefficients. Red regions indicate positive correlations, blue regions indicate negative correlations, and white regions indicate negligible correlations. Only property pairs with at least 30 data points ($N \geq 30$) are included.}
  \label{correlation}
\end{figure}

Having established the physical consistency of the MD-derived properties and quantified their systematic deviations relative to experiments, we now turn to data-driven polymer property modeling using ML. The validated MD, experimental, and GC datasets provide a complementary foundation for learning shared structure--property relationships across multiple targets. In the following sections, we examine how multi-task and multi-fidelity ML strategies leverage these datasets to improve predictive performance, particularly in data-scarce regimes.

Figure~\ref{correlation} presents a Spearman rank correlation heatmap of the polymer properties included in the analysis. Only property pairs with at least 30 shared data points are retained to ensure statistical reliability. The heatmap reveals clear blocks of correlated behavior within subsets of properties, along with weaker but non-negligible cross-property associations, highlighting the potential benefit of learning shared representations across multiple targets.
Strong correlation patterns are observed among electronic and dielectric descriptors, including $\alpha$, $E_g$, $E_{\mathrm{HOMO}}$, $E_{\mathrm{LUMO}}$, $\mu$, $n$, $\varepsilon$, $\epsilon_r$, and $E_{\mathrm{total}}$, consistent with their shared dependence on electronic structure and polarization response. Mechanical properties ($E$, $G$, $K$, and $\nu$) also exhibit internally consistent correlations, reflecting their common sensitivity to stiffness and packing constraints in amorphous polymers. Thermal properties ($C_p$, $T_g$, $T_m$, $\alpha_{\mathrm{T}}$, and $\kappa$) show more heterogeneous correlations, indicating that while certain structural factors influence multiple thermal responses, each property retains distinct physical drivers.
The heatmap further shows measurable cross-domain correlations between structural descriptors ($\rho$ and $R_g$) and several thermal and mechanical properties, as well as selected electronic and dielectric quantities. These patterns suggest that polymer packing and conformational statistics encode shared information that can support prediction across different property classes, even when direct physical coupling is indirect.
\begin{table}[t!]
\centering
\caption{Property groupings used in this work. Physical groups correspond to domain-based
categorization, while correlation-based groups (G1--G4) are derived from Pearson correlation
analysis and used for multi-task learning.}
\label{table2}
\begin{tabular}{lll}
\hline
\textbf{Group type} & \textbf{Group name} & \textbf{Properties} \\
\hline

Physical & Electronic 
& $\alpha$, $E_g$, $\varepsilon$, $E_{\mathrm{total}}$, 
  $E_{\mathrm{HOMO}}$, $E_{\mathrm{LUMO}}$, $\mu$, $n$ \\

Physical & Thermal 
& $C_p$, $T_g$, $T_m$, $\alpha_{\mathrm{T}}$, $\kappa$ \\

Physical & Mechanical 
& $K$, $\nu$, $G$, $E$ \\

Physical & Permeability 
& $P_{\mathrm{He}}$, $P_{\mathrm{H_2}}$, $P_{\mathrm{CO_2}}$, 
  $P_{\mathrm{N_2}}$, $P_{\mathrm{O_2}}$, $P_{\mathrm{CH_4}}$ \\

Physical & Other 
& $\eta$, $D$, $\rho$, $R_g$ \\

\hline
Correlation & Group 1 (G1) 
& $\alpha$, $E_g$, $E_{\mathrm{HOMO}}$, $E_{\mathrm{LUMO}}$, $n$, $E_{\mathrm{total}}$ \\

Correlation & Group 2 (G2) 
& $C_p$, $E_{\mathrm{LUMO}}$, $E_g$, $R_g$, $\rho$, $T_g$ \\

Correlation & Group 3 (G3) 
& $T_g$, $T_m$, $\eta$ \\

Correlation & Group 4 (G4) 
& $P_{\mathrm{He}}$, $P_{\mathrm{H_2}}$, $P_{\mathrm{CO_2}}$, 
  $P_{\mathrm{N_2}}$, $P_{\mathrm{O_2}}$, $P_{\mathrm{CH_4}}$, $C_p$, $\kappa$ \\
\hline
\end{tabular}
\end{table}
Guided by these observations, two complementary grouping strategies are adopted for multi-task learning, as summarized in Table~\ref{table2}. First, physical groupings are defined based on property domains (electronic, thermal, mechanical, permeability, and other). Second, correlation-based groups (G1--G4) are constructed by clustering properties that exhibit stronger Spearman rank correlations in Fig.~\ref{correlation}. These grouping schemes enable a systematic evaluation of whether correlation-aware task selection provides additional benefits beyond conventional domain-based multi-task learning, which is examined in the following section.

\begin{figure}[t!]
  \centering
  \includegraphics[width=1.05\textwidth]{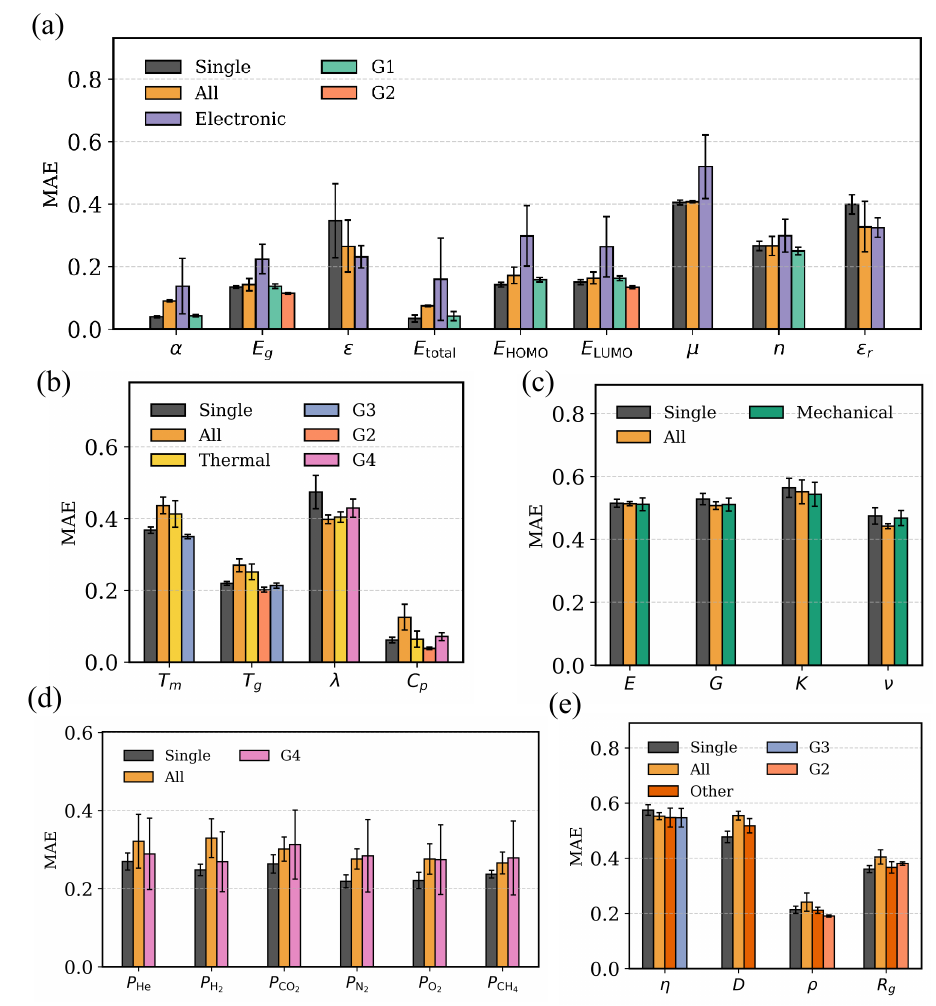}
  \caption{Comparison of scaled mean absolute error (MAE) for different polymer properties obtained from single-task and multi-task learning models. The multi-task models are trained using different sets of target properties, including all properties, property-class-based groups (thermal, mechanical, electronic, permeability, and others), and correlation-based groups (Groups 1–4).}
  \label{mae}
\end{figure}

We now turn to data-driven polymer property modeling using multi-task and multi-fidelity ML. Instead of training independent models for each target property, the proposed framework learns a shared molecular representation across multiple properties while using property-specific prediction heads. This design enables information sharing among related targets and allows lower-fidelity data to support representation learning, while higher-fidelity data provides stronger supervision where available. The following results evaluate the impact of task grouping and joint learning on predictive accuracy, using single-task models as a baseline.

Figure~\ref{mae} compares the scaled MAE obtained from single-task and multi-task ML models across different property groups and task configurations. In Fig.~\ref{mae}(a), electronic and dielectric properties generally show reduced error under multi-task learning, particularly when correlation-based groupings (G1 and G2) are used. In contrast, training a single model across all properties does not consistently improve performance, suggesting that indiscriminate task aggregation can yield limited gains or negative transfer for some electronic targets.
\begin{figure}[t!]
  \centering
  \includegraphics[width=1.0\textwidth]{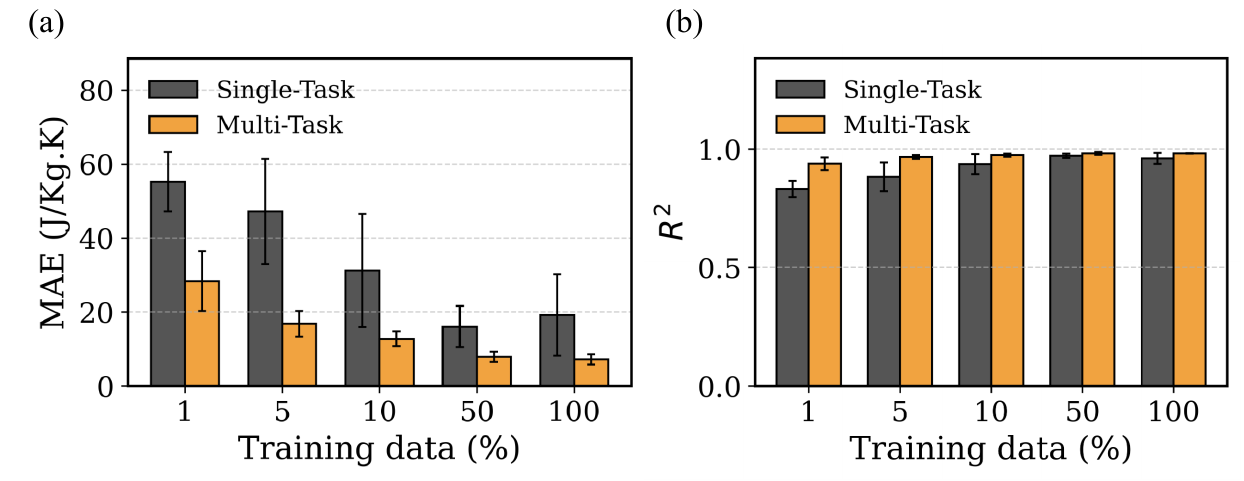}
  \caption{Comparison of $C_p$ prediction performance between single-task and multi-task models using different training data sizes.
(a) Mean absolute error (MAE).
(b) Coefficient of determination ($R^2$).}
  \label{sample}
\end{figure}
Thermal properties (Fig.~\ref{mae}(b)) exhibit a more mixed response. Multi-task learning improves prediction accuracy for $C_p$, whereas only modest or negligible changes are observed for $T_m$ and $T_g$. This behavior is consistent with the partially distinct physical mechanisms underlying these thermal properties. In several cases, correlation-based groupings perform comparably to, or even better than, purely domain-based groupings, suggesting that statistically informed task selection can be beneficial.
Mechanical properties ($E$, $G$, $K$, and $\nu$) show relatively small differences between single-task and multi-task models (Fig.~\ref{mae}(c)). This indicates that the available MD data already provide adequate supervision for these targets, limiting the additional benefit of task coupling. Transport and structural properties (Fig.~\ref{mae}(e)) display selective improvements under multi-task learning, particularly for $\rho$ and $R_g$, where shared structural information may contribute to improved predictions.
Overall, Fig.~\ref{mae} shows that multi-task learning can reduce prediction error when task groupings reflect meaningful statistical or physical relationships. However, the benefits are property-dependent, and joint learning may offer only limited improvement when unrelated properties are combined. These observations support the use of correlation-aware task grouping and provide a basis for the multi-fidelity analysis presented in the following section.

Figure~\ref{sample} highlights the robustness of the multi-task learning framework relative to single-task models under varying training data availability. The analysis examines the effect of progressively reducing the fraction of training data while keeping the test set fixed. Training fractions of 100\%, 50\%, 10\%, 5\%, and 1\% are considered, corresponding to approximately 10,000, 5,000, 1,000, 500, and 100 training data points, respectively.

When the full training dataset is used, single-task and multi-task models exhibit comparable performance in terms of both MAE and $R^2$, with the multi-task model showing slightly lower MAE and marginally higher $R^2$ values (Fig.~\ref{sample}). This indicates that, in the data-rich regime, both models are capable of learning accurate structure--property relationships for $C_p$.
As the training data size decreases, a consistent performance gap emerges across both error and correlation metrics. The multi-task model maintains lower MAE and higher $R^2$ values than the single-task model at all reduced training fractions. The difference becomes more pronounced at lower data availability, where the single-task model shows a clear degradation in predictive accuracy, while the multi-task model retains comparatively stable performance. This behavior is observed consistently in both MAE (Fig.~\ref{sample}(a)) and $R^2$ (Fig.~\ref{sample}(b)), indicating improved robustness of the multi-task model under limited training data.
These results indicate that multi-task learning improves data efficiency by leveraging shared information across related tasks, leading to better generalization when training data are scarce. The observed trends support the use of multi-task learning for polymer property prediction in scenarios where high-fidelity experimental data are limited.

\begin{figure}[t!]
  \centering
  \includegraphics[width=1.0\textwidth]{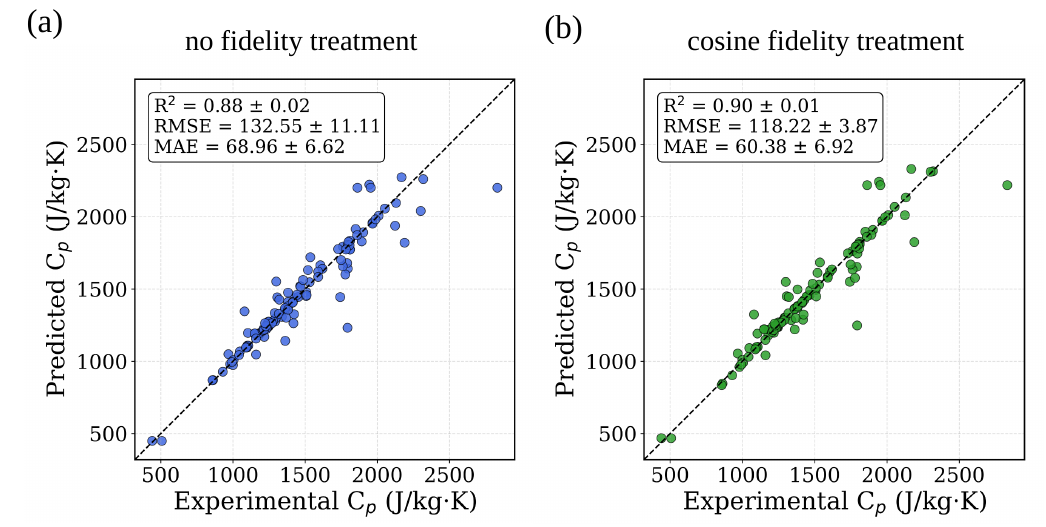}
  \caption{Parity plots comparing ML-predicted and experimental $C_p$.
(a) ML model trained with equal fidelity assigned to experimental and GC data.
(b) ML model trained with higher fidelity assigned to experimental data and lower fidelity assigned to GC data using a cosine weighting schedule.}
  \label{fidelity}
\end{figure}

Figure~\ref{fidelity} examines the effect of data fidelity treatment on the predictive performance of single-task $C_p$ models trained using combined GC and experimental datasets. The analysis compares parity behavior obtained with equal-fidelity training and with a fidelity-aware weighting strategy. In Fig.~\ref{fidelity}(a), equal fidelity weights are assigned to both datasets throughout training. This approach yields good overall agreement with experiments, with an $R^2$ value of $0.88 \pm 0.02$ and an MAE of $68.96 \pm 6.62$~J/kg.K.
Figure~\ref{fidelity}(b) shows results obtained using a fidelity-aware training strategy in which GC and experimental data are initially assigned equal weights, followed by a gradual increase in the fidelity weight of experimental data using a cosine schedule. This approach improves predictive accuracy, yielding an $R^2$ value of $0.90 \pm 0.01$ and an MAE of $60.38 \pm 6.92$~J/kg.K. Relative to the equal-fidelity case, this corresponds to a reduction of approximately 12\% in MAE. The corresponding parity plot exhibits tighter clustering around the parity line and reduced dispersion across the full $C_p$ range.
These results indicate that explicitly accounting for data fidelity during training enables the model to exploit the broad coverage of GC data for representation learning while progressively emphasizing experimental data to refine quantitative accuracy. The observed improvement with cosine fidelity weighting supports the use of fidelity-aware learning strategies when combining heterogeneous data sources for polymer property prediction.

\begin{table*}[t!]
\centering
\caption{Summary statistics of predicted polymer properties for the PolyInfo database
(13000 real polymers) and the PI1M  database (1 million virtual polymers).}
\label{table3}
\begin{tabular}{lccc ccc}
\hline
\textbf{Property} 
& \multicolumn{3}{c}{\textbf{PolyInfo Database}} 
& \multicolumn{3}{c}{\textbf{PI1M Database}} \\
\cmidrule(lr){2-4}\cmidrule(lr){5-7}
& Min & Median & Max & Min & Median & Max \\
\hline

\multicolumn{7}{l}{\textbf{Thermal}} \\
$T_m$ (K)                          & 207.87	& 509.66	& 880.45 & 203.78	& 447.25	& 918.02 \\
$T_g$ (K)                           & 143.58	& 394.01	& 702.26 & 113.38	& 338.99	& 691.51 \\
$\alpha_{\mathrm{T}}$ (m$^2$/s)  & 5.0$\times10^{-8}$ & 1.8$\times10^{-7}$ & 5.6$\times10^{-7}$ & 2.5$\times10^{-8}$ & 1.8$\times10^{-7}$ & 6.0$\times10^{-7}$\\
$\kappa$ (W/m.K)             & 0.11	& 0.26	& 0.49 & 0.14	& 0.26	& 0.83 \\
$C_p$ (J/kg.K)       & 437.43	& 1257.86	& 2560.19 & 551.00	& 1304.21	& 2796.52 \\[4pt]

\multicolumn{7}{l}{\textbf{Mechanical}} \\
$E$ (GPa)                          & 1.46	& 5.26	& 10.77 & 0.82	& 5.18	& 10.16 \\
$G$ (GPa)                          & 0.59	& 2.01	& 4.08  & 0.43	& 1.95	& 3.57\\
$K$ (GPa)                         & 1.20	& 4.90	& 8.61 & 1.24	& 4.84	& 9.95 \\
$\nu$                             & 0.08	& 0.31	& 0.43 & 0.08	& 0.32	& 0.51  \\[4pt]

\multicolumn{7}{l}{\textbf{Transport}} \\
$\eta$ (Pa$\cdot$s)                & 7.1$\times10^{-3}$ & 1.1$\times10^{-2}$ & 2.9$\times10^{-2}$ & 6.4$\times10^{-3}$ & 1.3$\times10^{-2}$ & 2.9$\times10^{-2}$ \\
$D$ (cm$^2$/s)              & 2.7$\times10^{-8}$ & 3.0$\times10^{-4}$ & 9.0$\times10^{-2}$ & 1.3$\times10^{-8}$ & 1.1$\times10^{-3}$ & 2.9$\times10^{-1}$ \\[4pt]

\multicolumn{7}{l}{\textbf{Gas Permeability}} \\
$P_{\mathrm{He}}$ (Barrer)         & 0.55	& 14.93	& 26495.19 & 1.14	& 21.80	& 37246.60 \\
$P_{\mathrm{H_2}}$ (Barrer)         & 0.56	& 17.83	& 38526.12  & 0.56	& 20.19	& 75825.66 \\
$P_{\mathrm{CO_2}}$ (Barrer)        & 2.3$\times10^{-4}$	& 9.81	& 61187.62 & 4.3$\times10^{-3}$	& 14.80	& 54059.37 \\
$P_{\mathrm{N_2}}$ (Barrer)         & 5.7$\times10^{-4}$	& 0.46	& 18698.51  & 1.5$\times10^{-4}$	& 0.86	& 40807.90\\
$P_{\mathrm{O_2}}$ (Barrer)         & 9.1$\times10^{-4}$	& 1.64	& 23378.65 & 6.4$\times10^{-4}$	& 2.17	& 30797.60 \\
$P_{\mathrm{CH_4}}$ (Barrer)        &  1.7$\times10^{-3}$	& 0.53	& 24869.14 & 5.6$\times10^{-4}$	& 1.85	& 58958.58 \\[4pt]

\multicolumn{7}{l}{\textbf{Electronic / Optical}} \\
$\alpha$ (a.u.)                    & 2.10	& 45.39	& 93.94  & 0.25	& 40.04	& 89.77\\
$E_{\mathrm{HOMO}}$ (eV)           & -12.96	& -8.67	& -6.74 & -11.98	& -9.37	& -8.09 \\
$E_{\mathrm{LUMO}}$ (eV)           & -1.84	& 0.57	& 3.24  & -2.97	& 1.03	& 3.17\\
$E_g$ (eV)                         & 6.57	& 9.15	& 15.48 & 7.55	& 10.91	& 16.70 \\
$\mu$ (Debye)                      & 0.02	& 2.90	& 6.92  & 0.01	& 2.97	& 8.24\\
$E_{\mathrm{total}}$ (eV)          & -3.0$\times10^{6}$	& -8.9$\times10^{5}$	& -4.7$\times10^{4}$ & -4.3$\times10^{6}$	& -8.3$\times10^{5}$	& -4.1$\times10^{4}$ \\
$n$                                & 1.26	& 1.55	& 1.74 & 1.25	& 1.52	& 1.91 \\
$\varepsilon$                      & 2.08	& 3.58	& 16.19 & 2.13	& 3.90	& 21.82 \\
$\epsilon_r$                       & 13.77	& 24.88	& 78.64  & 13.54	& 25.65	& 86.77 \\[4pt]

\multicolumn{7}{l}{\textbf{Structural}} \\
$R_g$ (\AA)                        & 9.21	& 19.49	& 42.95 & 8.50	& 18.06	& 43.35 \\
$\rho$ (g/cm$^{3}$)               & 0.69	& 1.24	& 2.79 & 0.85	& 1.20	& 2.80 \\

\hline
\end{tabular}
\end{table*}

\begin{figure}[t!]
  \centering
  \includegraphics[width=0.9\textwidth]{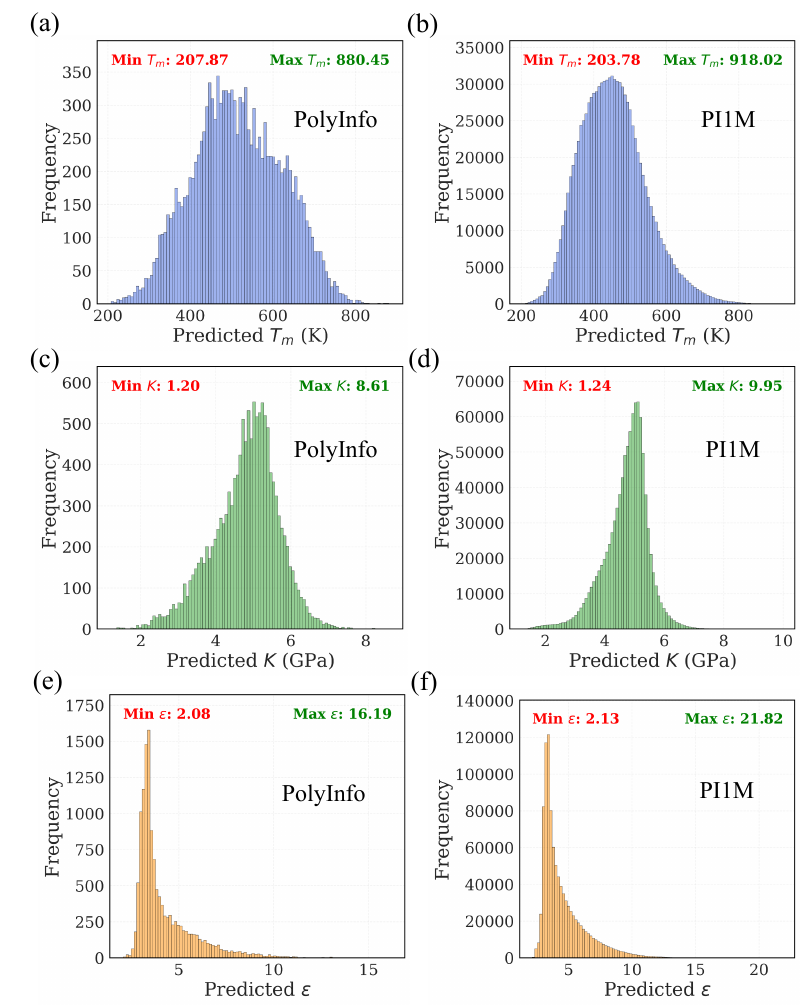}
  \caption{Predicted distributions of representative polymer properties obtained from the ML model for the PolyInfo and PI1M databases.
(a,b) Histograms of predicted $T_m$ for polymers in the PolyInfo and PI1M databases, respectively.
(c,d) Histograms of predicted  $K$ for PolyInfo and PI1M polymers.
(e,f) Histograms of predicted $\varepsilon$ for PolyInfo and PI1M polymers.
Minimum and maximum predicted values for each property are indicated in the corresponding panels.}
  \label{prediction}
\end{figure}

Having established the predictive accuracy and robustness of the proposed ML models, we next apply them to large-scale polymer property prediction. Large-scale property prediction is performed for polymers in the PolyInfo database, comprising approximately 13,000 real polymers, and the PI1M~\cite{ma2020pi1m} database, containing one million virtual polymers. In total, predictions were generated for 28 polymer properties, corresponding to approximately $3.6 \times 10^5$ property values for the PolyInfo dataset and $2.8 \times 10^7$ property values for the PI1M virtual polymer library. For each target property, the model configuration that achieved the best validation performance in the preceding analyses is selected. Depending on the property, this corresponds to either a single-task model or a multi-task model with an appropriate task grouping.
Table~\ref{table3} summarizes the minimum, median, and maximum predicted values for all properties across both datasets. For the PolyInfo database, the predicted ranges for thermal properties such as $T_m$, $T_g$, $\kappa$, and $C_p$ are consistent with experimentally reported values, with median predictions lying within physically expected regimes. Mechanical properties ($E$, $G$, $K$, and $\nu$) exhibit relatively narrow distributions, reflecting the constrained mechanical response typical of amorphous polymers. Electronic and optical properties display broader ranges, particularly for $\alpha$, $\mu$, and $\varepsilon$, consistent with their sensitivity to chemical composition and electronic structure. Structural properties ($R_g$ and $\rho$) show limited variation, reflecting packing constraints in polymer melts.
For the PI1M database, the predicted medians for most properties remain comparable to those observed for PolyInfo, while the overall property ranges are generally broader. This widening reflects the substantially larger and more diverse chemical space represented in the virtual polymer set. Importantly, predicted values across all properties remain within physically reasonable bounds, indicating stable model behavior when extrapolated to a larger design space.

Figure~\ref{prediction} presents representative predicted distributions for selected properties, including $T_m$, $K$, and $\varepsilon$, for both PolyInfo and PI1M polymers. The PolyInfo distributions show a single dominant peak with a spread comparable to that observed in experimental datasets, whereas the PI1M distributions exhibit smoother profiles and extended tails due to the larger sample size. The minimum and maximum values shown are consistent with the summary statistics reported in Table~\ref{table3} and further illustrate the broader coverage achieved for the virtual polymer library.
Together, the results in Table~\ref{table3} and Fig.~\ref{prediction} indicate that the trained models generate physically consistent property distributions at scale while preserving trends observed in experimentally characterized polymers. This capability enables systematic screening and comparative analysis of large polymer libraries across multiple property dimensions.

Taken together, the results indicate that combining physics-based simulations, experimental data, and ML within a unified framework enables consistent and scalable prediction of multiple polymer properties. Validation against experimental measurements provides confidence in the simulation-derived datasets, while correlation-aware multi-task learning and fidelity-weighted training improve predictive performance, particularly when experimental data are limited. The large-scale predictions for both real and virtual polymer libraries demonstrate the potential of the proposed framework for systematic evaluation of polymer properties across diverse chemical spaces.

\section{Conclusion}
In this work, an integrated polymer informatics framework is presented that combines physics-based data generation with data-driven modeling for the prediction of polymer properties. The ADEPT workflow enables automated evaluation of polymer properties from repeat-unit SMILES, using MD and DFT to compute thermal, mechanical, transport, structural, and electronic descriptors consistently. When combined with experimentally curated measurements and GC estimates, this approach results in a heterogeneous dataset of approximately 62,000 property values spanning multiple property classes and data fidelity levels.

Using this dataset, inter-property correlations are quantified to assess the suitability of joint learning across targets. The results show that the effectiveness of multi-task learning depends on task-relatedness, with correlation-informed groupings providing more consistent improvements than uniform task aggregation. The analysis further indicates that multi-task learning improves data efficiency, with performance differences between single-task and multi-task models becoming more pronounced as the available training data are reduced. In addition, fidelity-aware training strategies improve predictive accuracy when combining experimental and computational data by balancing broad coverage from lower-fidelity sources with stronger supervision from experimental measurements.

The trained models are subsequently applied to large-scale property prediction for polymers in the PolyInfo database and the PI1M virtual polymer library. The resulting property distributions remain physically reasonable and reflect trends observed in experimentally characterized polymers, supporting the use of the models for systematic screening across extended chemical spaces.

Overall, the results demonstrate that integrating simulation-derived data, experimental measurements, and ML within a unified framework provides a practical approach for multi-property polymer prediction. The proposed methodology supports the scalable evaluation of both known and virtual polymers and provides a structured framework for polymer screening and analysis in settings where data availability and fidelity vary across properties.

\section{Methods}
\subsection{ADEPT Workflow: Polymer Simulation and Property Generation}
\subsubsection{Polymer Structure Generation from SMILES}

Polymer repeat units were defined represented by SMILES of monomers, with polymerization sites specified using isotope-labeled terminal hydrogens denoted as "*" to identify head and tail atoms. Three-dimensional monomer geometries were generated using the ETKDG algorithm~\cite{riniker2015better} and imported into the PySIMM framework for polymer construction~\cite{fortunato2017pysimm}. Polymer chains were built using a random-walk polymerization algorithm~\cite{theodorou1985detailed}, with the chain length selected to yield approximately 600 atoms per polymer chain, ensuring comparable molecular weights across different polymers. During chain growth, terminal hydrogens were removed, and the final polymer chains were capped at both ends with methyl (–CH$_3$) groups.
All polymer structures were parameterized using the General AMBER Force Field (GAFF2) force field~\cite{wang2004development}. GAFF2 was used for all MD simulations due to its broad chemical coverage and established parameterization for polymers; a detailed force-field sensitivity analysis for $C_p$ calculations is reported in our previous work and is not repeated here\cite{alosious2026transfer}. Atomic partial charges were assigned using either the Restrained Electrostatic Potential (RESP) method~\cite{bayly1993well} or Gasteiger charges~\cite{gasteiger1980iterative}. RESP charges were obtained from gas-phase quantum chemical calculations performed on optimized monomer geometries, whereas Gasteiger charges were used for computationally efficient simulations when dielectric properties were not required. After the force-field assignment, the polymer chains were energy-minimized and replicated to generate amorphous configurations consisting of six chains. These configurations were initially packed at a $\rho$ of approximately 0.01~g/cm$^{3}$. The resulting amorphous polymer structures were written in formats compatible with LAMMPS~\cite{plimpton1995fast} for subsequent MD simulations.

\subsubsection{Amorphous Polymer Generation and Equilibration}

Following polymer chain construction and force-field assignment, bulk amorphous polymer structures were generated and equilibrated using MD simulations in LAMMPS. Multiple polymer chains were randomly packed into a three-dimensional simulation cell at a low initial $\rho$ and equilibrated using a multi-stage protocol comprising an initial relaxation and annealing.

\noindent
\textbf{Initial relaxation:}  
Electrostatic interactions were initially disabled, and Lennard-Jones (LJ) interactions were truncated at a cutoff distance of 0.3~nm to reduce large forces arising from unfavorable contacts in the randomly packed configuration. The system was first equilibrated under an NPT ensemble at 100~K for 2~ps using a 0.1~fs time step. The temperature was then increased from 100~K to 1000~K over 1~ns under NVT conditions. This was followed by equilibration at 1000~K and 0.1~atm for 50~ps under NPT conditions, and a subsequent 1~ns NPT simulation during which the pressure was gradually increased from 0.1~atm to 500~atm using a 1~fs time step. SHAKE~\cite{ryckaert1977numerical} constraints were applied throughout this stage to constrain covalent bond lengths and ensure numerical stability.

\noindent
\textbf{Annealing:}  
Electrostatic interactions were re-enabled using the particle--particle--particle--mesh (PPPM) Ewald summation method~\cite{hockney2021computer}, and the LJ cutoff distance was increased to 0.800~nm. The system was equilibrated at 1000~K and 1~atm under NPT conditions for 2~ps using a 0.1~fs time step. The system was then cooled to 300~K at a rate of 140~K/ns while maintaining NPT conditions and SHAKE constraints. A final NPT simulation was performed at 300~K and 1~atm for 8~ns using a 1~fs time step to obtain a stable amorphous polymer configuration for subsequent property calculations.

\subsubsection{Property Calculations from Molecular Dynamics Simulations}

Unless otherwise specified, all polymer properties were computed from equilibrated amorphous configurations obtained after the annealing protocol described above. Production simulations were performed under NPT conditions at $300~\mathrm{K}$ and $1~\mathrm{atm}$, and property values were obtained by averaging over statistically converged time intervals. Electronic properties were evaluated at the monomer level using DFT calculations. The following subsections describe the methodologies used for calculating structural, thermodynamic,  transport,  mechanical,   electronic, and dielectric properties.

\subsubsection{Structural Properties}

\textbf{Density ($\rho$).} The mass density  was computed as the ratio of the total system mass to the time-averaged simulation cell volume,
\begin{equation}
\rho = \frac{m}{\langle V \rangle}.
\end{equation}
\textbf{Radius of gyration ($R_g$).} The radius of gyration  was calculated for each polymer chain according to
\begin{equation}
R_g = \sqrt{\frac{1}{N}\sum_{k=1}^{N}\left(\mathbf r_k - \mathbf r_{\mathrm{cm}}\right)^2},
\end{equation}
where $N$ is the number of atoms in the chain, $\mathbf r_k$ is the position of atom $k$, and $\mathbf r_{\mathrm{cm}}$ is the center-of-mass position of the chain. Reported $R_g$ values correspond to averages taken over all polymer chains and over statistically converged time intervals.

\subsubsection{Thermal and Thermodynamic Properties}

\textbf{Glass transition temperature ($T_g$).} The glass transition temperature was determined from density--temperature curves obtained during controlled NPT cooling simulations. Linear regressions were performed separately for the low-temperature and high-temperature regions, and $T_g$ was identified as the intersection of the corresponding fitted lines,
\begin{equation}
T_g = \frac{c_{\mathrm{high}} - c_{\mathrm{low}}}{m_{\mathrm{low}} - m_{\mathrm{high}}},
\end{equation}
where $m$ and $c$ denote the slopes and intercepts of the respective linear fits. To reduce sensitivity to the choice of fitting window, multiple fits were performed using different numbers of data points in each temperature regime. The final $T_g$ value was taken as the mean of the resulting intersection temperatures. Any fitted $T_g$ value falling outside the initial temperature-scanning range was discarded.
For each polymer, the temperature-scanning range was defined as a bounded interval around an initial estimate of $T_g$. This initial estimate was obtained from a separate multilayer perceptron model trained on experimentally reported $T_g$ values. The MD temperature scan was then performed over a restricted window of $\pm$150 K around the estimated $T_g$, rather than across a broad global temperature interval.
 This procedure limits the required simulation window while ensuring adequate sampling of both glassy and rubbery regimes.\\
\textbf{Specific heat capacity ($C_p$).} The constant-pressure specific heat capacity was evaluated using both equilibrium MD (EMD) and non-equilibrium MD (NEMD) simulations under NPT conditions. The use of both approaches allows assessment of consistency while balancing accuracy and computational cost.

In the equilibrium approach, after sufficient equilibration, production simulations were performed at 300~K and 1~atm to sample thermodynamic fluctuations. The $C_p$ was computed from enthalpy fluctuations using the fluctuation--dissipation relation~\cite{allen2017computer},
\begin{equation}
C_p = \frac{\langle H^2 \rangle - \langle H \rangle^2}{k_B T^2\, m},
\end{equation}
where $H$ is the instantaneous system enthalpy, $T$ is the absolute temperature, $k_B$ is Boltzmann’s constant, and $m$ is the total system mass. Angle brackets denote ensemble averages over the equilibrated trajectory. This approach can be evaluated concurrently with other equilibrium properties, making it computationally efficient for large-scale screening studies.

In addition, a non-equilibrium approach based on the enthalpy--temperature relationship was employed. The system was first equilibrated at 290~K and 1~atm, followed by a controlled temperature ramp from 290~K to 310~K under NPT conditions. Average enthalpy values were recorded as a function of temperature, and $C_p$ was obtained from the slope of a linear fit to the enthalpy--temperature curve~\cite{bhowmik2019calculation},
\begin{equation}
C_p = \left( \frac{dH}{dT} \right)_P.
\end{equation}
This non-equilibrium approach generally provides more stable estimates of $C_p$, but requires separate temperature-ramping simulations and therefore incurs additional computational cost; accordingly, all $C_p$ values reported in this work were obtained using the NEMD method. A detailed comparison between EMD and NEMD approaches for polymer $C_p$, including quantitative error analysis, is provided in our previous work~\cite{alosious2026transfer}, where NEMD was shown to offer improved numerical stability and was therefore adopted in the present study.\\
\textbf{Thermal expansion coefficient ($\alpha_V$).} The volumetric thermal expansion coefficient was computed from temperature-dependent volume data according to
\begin{equation}
\alpha_V = \frac{1}{\langle V \rangle}\left(\frac{\mathrm d \langle V \rangle}{\mathrm d T}\right)_P,
\end{equation}
and the linear thermal expansion coefficient was obtained as $\alpha_L = \alpha_V/3$.\\
\textbf{Thermal conductivity ($\kappa$).} Thermal conductivity was computed using a NEMD approach in which a steady-state temperature gradient was imposed along a designated transport direction~\cite{ma2022machine}. After equilibration, localized hot and cold regions were maintained at fixed temperatures to induce a constant heat flux across the system. The simulation cell was divided into spatial bins along the transport direction, and the temperature profile was constructed from time-averaged kinetic temperatures within each bin.

The temperature gradient was extracted from the central conduction region by fitting the linear portion of the steady-state temperature profile, excluding bins adjacent to the thermostatted regions to minimize boundary effects. The heat flux was obtained from the cumulative energy exchanged with the thermal reservoirs and normalized by the cross-sectional area perpendicular to the transport direction. $\kappa$ was then evaluated using Fourier’s law,
\begin{equation}
\kappa = -\,\frac{J_q}{\mathrm{d}T/\mathrm{d}x},
\end{equation}
where $J_q$ is the steady-state heat flux and $\mathrm{d}T/\mathrm{d}x$ is the temperature gradient along the transport direction.

To ensure reliable estimation of $\kappa$, only simulations exhibiting a linear increase in cumulative exchanged energy with time and a well-defined linear temperature gradient in the central region were retained for analysis. Reported $\kappa$ values correspond to block-averaged means computed over the steady-state portion of the trajectory, with uncertainties estimated from the variance across independent time blocks.\\
\textbf{Thermal diffusivity ($\alpha_T$).} The thermal diffusivity was computed from $\kappa$, $\rho$, and $C_p$ as
\begin{equation}
\alpha_T = \frac{\kappa}{\rho\,C_p},
\end{equation}
using $\rho$ and $C_p$ obtained from NPT production simulations performed at $300~\mathrm{K}$ and $1~\mathrm{atm}$.

\subsubsection{Transport Properties}

\textbf{Self-diffusion coefficient ($D$).} The self-diffusion coefficient was computed from the mean-square displacement using the Einstein relation,
\begin{equation}
D = \lim_{t \to \infty} \frac{1}{6t}\left\langle \lvert \mathbf r(t) - \mathbf r(0) \rvert^2 \right\rangle,
\end{equation}
where $\mathbf r(t)$ denotes the position of a particle at time $t$, and angle brackets indicate an ensemble average over particles and time origins.\\
\textbf{Viscosity ($\eta$).} The shear viscosity was obtained from EMD trajectories using the Green--Kubo formalism,
\begin{equation}
\eta = \frac{\langle V \rangle}{k_B T}\int_0^\infty
\left\langle P_{\alpha\beta}(0)\,P_{\alpha\beta}(t) \right\rangle \mathrm dt,
\qquad \alpha \neq \beta,
\end{equation}
where $P_{\alpha\beta}$ are the off-diagonal components of the pressure tensor. Ensemble averages were taken over independent time origins and Cartesian shear components.

\subsubsection{Mechanical properties}

The elastic properties of amorphous polymers were calculated using a finite-deformation stress-strain approach. Each equilibrated polymer cell was first converted from a cubic to a triclinic simulation box to enable shear as well as normal deformations. Symmetric finite strains of magnitude $\pm 2\%$ were applied along the six independent strain modes, comprising three normal and three shear components. Following each deformation, the system was simulated at $300~\mathrm{K}$ using NVE integration coupled with a Langevin thermostat, consisting of a $250~\mathrm{fs}$ brief equilibration followed by a $75~\mathrm{fs}$ sampling segment with a $0.25~\mathrm{fs}$ timestep, and the time-averaged stress tensor was recorded.

Within linear elasticity, the stress--strain relationship is given by
\begin{equation}
\sigma_i = \sum_{j=1}^{6} C_{ij}\,\varepsilon_j,
\end{equation}
where $\sigma_i$ and $\varepsilon_j$ denote the stress and strain components in Voigt notation. The elastic stiffness coefficients $C_{ij}$ were obtained from central finite differences of the stress response with respect to the applied strain, and the full $6 \times 6$ stiffness matrix was constructed.

For isotropic amorphous systems, the bulk modulus ($K$) and shear modulus ($G$) were computed using the Voigt--Reuss--Hill averaging scheme applied to the stiffness and compliance tensors. Young’s modulus ($E$) and Poisson’s ratio ($\nu$) were then derived from $K$ and $G$ using standard isotropic relations~\cite{allen2017computer},
\begin{equation}
E = \frac{9KG}{3K+G},
\end{equation}
\begin{equation}
\nu = \frac{3K-2G}{2(3K+G)}.
\end{equation}

To ensure statistical reliability, for each simulation, final mean values of elastic properties were averaged over 100 configurations evenly sampled from an $8~\mathrm{ns}$ production trajectory. Reported values correspond to ensemble averages, with coefficients of variation (defined as the ratio of standard deviation to the mean) not exceeding 0.33.

\subsubsection{Electronic and dielectric properties from DFT}

Electronic descriptors were computed at the monomer level using DFT to obtain the static dipole polarizability ($\alpha$), highest occupied molecular orbital energy ($E_{\mathrm{HOMO}}$), lowest unoccupied molecular orbital energy ($E_{\mathrm{LUMO}}$), electronic band gap ($E_g$), total electronic energy ($E_{\mathrm{total}}$), and molecular dipole moment ($\mu$). Monomer-based calculations were employed to enable scalable evaluation across large polymer libraries while retaining sensitivity to repeat-unit chemistry.

Monomer geometries were generated from repeat-unit SMILES using RDKit, where polymerization placeholders (\texttt{*}) were replaced with tritium tags (\([\!^{3}\mathrm H]\)) to preserve valence and identify connection sites \cite{landrum2013rdkit}. Initial three-dimensional coordinates were generated using the ETKDGv2 algorithm with explicit hydrogens added prior to export in XYZ format \cite{riniker2015better}.

All electronic-structure calculations were performed using Psi4 \cite{smith2020psi4}. To obtain stable and physically reasonable geometries, a three-stage optimization protocol was applied: (i) HF/STO-3G pre-optimization, (ii) refinement at HF/6-31G, and (iii) final optimization using the range-separated, dispersion-corrected $\omega$B97M-D3BJ functional with the 6-311+G(2d,p) basis set \cite{mardirossian2016omegab97m}. Geometry and self-consistent field convergence thresholds were progressively tightened at each stage.

The static dipole polarizability was evaluated at the optimized geometry using a finite-field approach. Single-point DFT calculations were performed under uniform electric fields of magnitude $\pm \delta$, with $\delta = 10^{-4}$ atomic units, applied independently along each Cartesian direction. The polarizability tensor was obtained from central finite differences of the induced dipole moments,
\begin{equation}
\alpha_{ij} \approx -\frac{\mu_i^{(+)} - \mu_i^{(-)}}{2\delta},
\end{equation}
and the isotropic polarizability was computed as $\bar{\alpha} = \tfrac{1}{3}\mathrm{Tr}\,\boldsymbol{\alpha}$ and reported in $\text{\AA}^3$ \cite{buckingham1967permanent}.

Single-point electronic properties were then computed at the optimized geometries using $\omega$B97M-D3BJ with density fitting and tight self-consistent field convergence. HOMO and LUMO energies were extracted from the converged orbital spectrum and reported in eV, and the electronic band gap was computed as $E_g = E_{\mathrm{LUMO}} - E_{\mathrm{HOMO}}$. Total electronic energies were reported in kcal/mol, and dipole moments were reported in Debye. For iodine-containing monomers, a mixed-basis treatment was employed in which LanL2DZ was applied to iodine while all other atoms used 6-311G(d,p) \cite{hay1985ab}.

The electronic contribution to the dielectric response was estimated by first computing the refractive index ($n$) from the DFT-derived isotropic polarizability using the Lorentz–Lorenz relation \cite{lorentz1916theory},
\begin{equation}
\frac{n^2 - 1}{n^2 + 2} = \frac{4\pi}{3} N \bar{\alpha},
\end{equation}
where $N$ is the number density and $\bar{\alpha}$ is the isotropic polarizability. The electronic dielectric constant was then obtained as $\varepsilon_{\mathrm{el}} = n^2$.

The orientational, or dipolar, contribution to the static dielectric constant ($\varepsilon_{\mathrm{dip}}$) was obtained from MD simulations using dipole moment fluctuations,
\begin{equation}
\varepsilon_{\mathrm{dip}} = 1 + \frac{\langle M^2 \rangle - \langle \mathbf M \rangle^2}{3\,\varepsilon_0\,k_B T\,\langle V \rangle},
\end{equation}
where  $\mathbf M$ is the total dipole moment of the simulation cell, $\varepsilon_0$ is the vacuum permittivity, $k_B$ is Boltzmann’s constant, $T$ is the absolute temperature, and $\langle V \rangle$ is the time-averaged simulation cell volume. Angle brackets denote ensemble averages taken over the equilibrated trajectory.

The total static dielectric constant ($\varepsilon$) was calculated as
\begin{equation}
\varepsilon = \varepsilon_{\mathrm{el}} + \varepsilon_{\mathrm{dip}} - 1,
\end{equation}
and the absolute permittivity was obtained as $\epsilon = \varepsilon\,\varepsilon_0$.

\subsection{PolyGraphMT Framework: Multi-Task Multi-Fidelity Learning}
\subsubsection{Loss function and normalization}

Model parameters were optimized by minimizing a composite loss function defined as
\begin{equation}
\mathcal{L} =
\sum_{p \in \mathcal{P}} \sum_{f \in \mathcal{F}_p}
\frac{w_{p,f}}{N_{p,f}}
\sum_{i=1}^{N_{p,f}}
\left( \hat{y}_{i,p} - y_{i,p}^{(f)} \right)^2,
\end{equation}
where $N_{p,f}$ denotes the number of available data points for property $p$ at fidelity level $f$, and $w_{p,f}$ is a weighting factor used to balance the contributions from different properties and data fidelities.

The normalization by $N_{p,f}$ ensures that properties or fidelity levels with larger datasets do not dominate the optimization objective. The weighting factors $w_{p,f}$ provide a mechanism to adjust the relative influence of experimental and computational data during training, reflecting their differing reliability without enforcing explicit bias correction.

\subsubsection{Training protocol and hyperparameter optimization}

Models were trained using mini-batch stochastic optimization on GPU-accelerated hardware. Data splits were defined at the SMILES level and kept fixed across all experiments to enable consistent comparison between single-task, multi-task, and multi-fidelity model configurations.
Hyperparameters, including learning rate, batch size, network depth, and regularization parameters, were optimized using Bayesian optimization implemented with Optuna \cite{akiba2019optuna}. Model selection was based on validation performance evaluated using mean absolute error (MAE), root mean square error (RMSE), or coefficient of determination ($R^2$), depending on the target property. For multi-task models, validation metrics were computed either as aggregated measures across tasks or with respect to a designated primary task, while the training objective itself remained unchanged.
All reported results were averaged over five random seeds. Variability across seeds was used to estimate uncertainty in model performance.

\backmatter

\section*{Acknowledgements}

This work was supported by the Lucy Family Institute for Data \& Society Postdoctoral Fellowship and National Science Foundation grants 2332270 and 2102592. The authors also acknowledge the Center for Research Computing at the University of Notre Dame for providing the necessary computing resources.

\section*{Declarations}

\bmhead{Competing interests}

The authors declare no competing interests.

\bmhead{Data availability}

The ADEPT workflow for automated polymer structure generation, MD simulations, and DFT calculations is available at \url{https://github.com/sobinalosious/ADEPT}. 
The multi-task and multi-fidelity ML models used for polymer property prediction are implemented in a separate repository at \url{https://github.com/sobinalosious/PolyGraphMT}. 
The processed datasets and scripts required to reproduce the reported results are provided in the corresponding repositories.

\bmhead{Ethics approval and consent to participate}
Not applicable
\bmhead{Consent for publication}
Not applicable
\bmhead{Materials availability}
Not applicable
\bmhead{Author contributions}

S.A. designed the study, developed the workflow, performed simulations, and implemented the machine learning models. Y.L., J.X., and R.Z. contributed to data generation and analysis. G.L. contributed to machine learning methodology and implementation. M.J. and T.L. supervised the research and contributed to the conceptual development of the study. All authors contributed to writing and reviewing the manuscript.

\bibliography{sn-bibliography}

\end{document}